\documentclass[a4paper, conference]{IEEEtran}
\usepackage{cite}

\usepackage{graphicx}
\usepackage{epstopdf}
\graphicspath{{./diagrams/}{./results/}}
\DeclareGraphicsExtensions{.eps}

\usepackage[cmex10]{amsmath}
\usepackage[lined, linesnumbered, ruled]{algorithm2e}

\usepackage[tight,footnotesize]{subfigure}
\begin{document}
%
\title{Adaptive TXOP Assignment for QoS Support of Video Traffic in IEEE 802.11e Networks}


%
\author{\IEEEauthorblockN{Mohammed A. Al-Maqri\IEEEauthorrefmark{1},
Mohamed Othman\IEEEauthorrefmark{1},
Borhanuddin Mohd Ali\IEEEauthorrefmark{7}, 
Zurina Mohd Hanapi\IEEEauthorrefmark{1}}

\IEEEauthorblockA{\IEEEauthorrefmark{1}Faculty of Computer Science and Information Technology,
\IEEEauthorrefmark{7}Faculty of Engineering\\ Universiti Putra Malaysia, 43400, Serdang, Selangor, Malaysia}
Email: mohdalmoqry@gmail.com, mothman@upm.edu.my }


\maketitle

\begin{abstract}
Quality of Service (QoS) is provided in IEEE 802.11e protocol by means of HCF Controlled Channel Access (HCCA) scheduler which is efficient for supporting Constant Bit Rate (CBR) applications. Numerous researches have been carried out to enhance the HCCA scheduler attempting to accommodate the needs of Variable Bit Rate (VBR) video traffics which probably demonstrates a non-deterministic profile during the time. This paper presents an adaptive TXOP assignment mechanism for supporting the transmission of the prerecorded video traffics over IEEE 802.11e wireless networks. The proposed mechanism uses a feedback about the size of the subsequent video frames of the uplink traffic to assist the Hybrid Coordinator (HC) accurately assign TXOP according to the fast changes in the VBR profile. The simulation results show that our mechanism reduces the delay experienced by VBR traffic streams comparable to HCCA scheduler due to the accurate assignment of the TXOP which preserve the channel time for data transmission.
\end{abstract}


%
\IEEEpeerreviewmaketitle

\section{Introduction}
\label{sec:Introduction}
Due to the wide spread of ubiquitous applications in the internet and the rapid growth of multimedia streams, providing  differentiated Quality of Service (QoS) support for such applications in Wireless Local Area Networks (WLANs) has become a very challenging task. IEEE802.11 \cite{IEEEStand1999} is the most commonly used technology in WLANs including two access modes, Distributed Coordination Function (DCF) and Point Coordination Function (PCF). The former deemed compulsory medium access method which serves best effort applications. Multimedia streams that require a certain QoS level are served during the polling-based  (i.e. PCF). However, it is not efficient enough to support high QoS requirement applications due to the fact that PCF only operates on Free-Contention period which may noticeably cause an increase in the transmission delay especially with high bursty traffics. Consequently, IEEE 802.11 Task Group e (TGe) has presented IEEE 802.11e protocol \cite{IEEEStandard2007} which introduces two Medium Access Control (MAC) modes, distributed and controlled. Enhanced Distributed Channel Access (EDCA) function which operates in a distributed manner to provide prioritized QoS and Hybrid Coordination Fucntion (HCF) Controlled Channel Access (HCCA) that introduces a polling mechanism which provides parameterized QoS for applications that require rigorous QoS requirements.

By 2014, about 91 percent of web traffic will be video streams \cite{DigitalMedia2012}. This fact motivates several researches to be carried on to improve the performance of WLANs in terms of provisioning QoS for such streams. Motion Picture Experts Group type 4 (MPEG--4/H.264) has become a prominent video the internet due to its scalability, error robustness and network-friendly features. MPEG--4 is available in various data bit rate to accommodate different network capacities. Delay-sensitive multimedia streams, such as scalable MPEG--4 are more adequate to be transmitted throughout HCCA as it was designated to minimize the overhead of messaging caused by the distributed approach of EDCA function. The hybrid coordinator (HC) in HCCA polls wireless stations periodically and allocates transmission opportunities (TXOP) to them. And yet, HCCA schedules traffics upon their QoS requirements negotiated in the first place, it is only efficient for constant bit rate applications such as CBR G.711 \cite{G7111988}, audio streams, and (H.261/MPEG-1) video \cite{MPEG11997}. HCCA is not convenient to deal with the fluctuation of the VBR traffic such as MPEG--4 video streams, where the packet size shows high variability during the time. This consequently leads to a remarkable increase in the end-to-end delay of the delivered traffics and degradation in the channel utilization.

HCCA scheduler computes TXOP so as to grant a QSTA an ample time to transmit its traffics during SI cycle. This calculation is based on the mean characteristics of the traffic which is not accurate, because of the deviation of VBR traffics from its mean characteristics,to clear the QSTA transmission queue at the end of SI. In this paper, we present an enhancement on the HCCA scheduling algorithm aiming to adapt to the fast fluctuation of VBR video traffics profile. Basically, the proposed mechanism computes the TXOP for a traffic based on knowledge about the actual frame size instead of assigning TXOP according to mean characteristics of the traffic which is unable to reflect the actual traffic. This mechanism makes use of the queue size field of QoS data frame in IEEE 802.11e MAC header to carry this information to the HC.

The rest of this paper is organized as follows. Section ~\ref{sec:relatedWorks} explains the reference HCCA mechanism and demonstrates its deficiency in supporting VBR Video streams. Section ~\ref{sec:ATAV} explains the proposed mechanism. The performance evaluation and discussion is presented in Section ~\ref{sec:evaluation}. Section ~\ref{sec:conclusion} concludes the study presented in this paper.

 
\section{Background and Related Works}
\label{sec:relatedWorks}
This section describes IEEE 802.11e HCCA scheduler along with some characteristics of  MPEG--4 VBR video traffic. The deficiency of HCCA in supporting VBR is explained. Some related works in enhancing its performance are also discussed.

\subsection{HCF Controlled Channel Access}
\begin{figure}[hbtp]
\centering
\includegraphics[width=\linewidth]{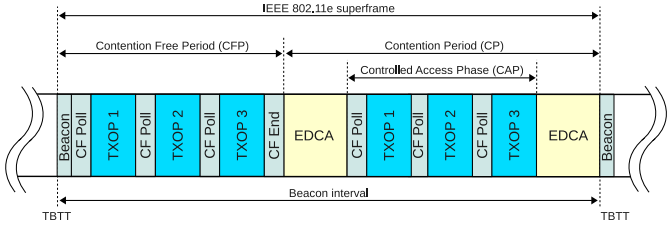}
\caption{Controlled Channel Access Mechanism in IEEE 802.11e HCCA}
\label{fig001}
\end{figure}

In IEEE 802.11e, a parameterized QoS is supported during HCCA using polling access method. A beacon is transmitted every Target Beacon Transmission Time (TBTT) comprising a superframe. Fig. \ref{fig001} demonstrate the superframe which includes Contention Free Period (CFP) followed by Contention Period (CP). HC uses both periods to deliver its data and polls the stations to transmit their uplink traffics. HCCA outperforms PCF of legacy IEEE802.11, in that it can be initiated in both CFP and CP in contrary to its ancestor, PCF, which only operates during  CFP. The HC may begin a Controlled Access Phase (CAP) at any time during the CP if the medium remains idle for a time equals to PCF Interframe Space (PIFS). This merit assists HCCA to considerably reduce the delay experienced by the traffics as they can be polled more frequently than that in PCF. When a station intends to initiate a data traffic, it issues a QoS reservation through a special QoS management action frame called ADDTS-Request contains a set of parameters that define the characteristics of the TS (TSPEC). The fields of the TSPEC and how the HC exploits them in the scheduling process is discussed in details in the next section.
\subsection{Reference Design of HCCA scheduler}
\label{sec:HCCA}
As mentioned earlier, the QoS-enabled station (QSTA) issues a QoS reservation through transmitting an ADDTS-Request frame. This frame carries information about the TS characteristics (TSPEC) which is required by HC for scheduling purpose. The mandatory fields of the TSPEC are described as follows:

\textit{Mean Data Rate} ($\rho$): the average data rate of the packets in units of bits per second.

\textit{Nominal MSDU Size} ($L$): the mean size of the MAC packet in units of bytes.

\textit{Maximum MSDU Size} ($M$): the maximum allowable size of the MAC packet in the TS in units of bytes.

\textit{Delay Bound ($D$):} the maximum allowed delay for a packet to be transmitted through the wireless medium in units of milliseconds.

\textit{Service Interval ($SI$):} the time interval between TXOPs of the station in units of milliseconds.

\textit{Physical Rate ($R$):}  the assumed physical bit rate of the wireless channel, in units of bits per second.

The HC which usually resides in the QoS-enabled Access Point (QAP) maintains theses TSPECs in so-called polling list. HC computes the duration of the time to be granted to each QSTA to transmit its traffics (TXOP). The admission of the TSs is governed by HC, using the Admission Control Unit (ACU). The HC reserves the right to accept or reject any TS so as to preserve the QoS of the previously admitted TSs. If HC accepts the traffic it will respond by an ADDTS-Response or a rejection message otherwise.

Upon receiving an ADDTS-Request from a QSTA, the HCCA scheduler goes through the following steps:
\subsubsection{SI Assignment}
\label{eq:SIassign}
The scheduler calculates $SI$ as the minimum of all Maximum Service Intervals ($MSI$) of all admitted traffic streams which is a submultiple of the beacon interval. The minimum $MSI$ for each QSTA is obtained from Equation~\eqref{eq:minSI}.
\begin{equation}
\label{eq:minSI}
MSI_{min} = min(MSI_{i}), \hspace{2mm} i =1,2,3, \cdotp\cdotp\cdotp, n
\end{equation}
where $n$ is the number of admitted QSTAs' traffic streams and $MSI_{i}$ is the maximum $SI$ of the $i^{th}$ stream.
The final $SI$ is computed so that it satisfies the condition in Equation~\eqref{eq:si}.
\begin{equation}
\label{eq:si}
SI = \frac{BeaconInterval}{x} \leq  MSI_{min}
\end{equation}
The denominator $x$ is an integer number that divides the beacon interval into a largest number equal or less than the $MSI_{min}$.

\subsubsection{TXOP Allocation}
HC allocates different TXOP to each admitted QSTA so as to transmit their data with respect to the negotiated QoS parameters of the TSPEC. This TXOP is calculated for each station as follows:

Firstly, for the $i^{th}$ QSTA, the scheduler calculates the number of MSDUs that may arrive at $\rho_{i}$ as in Equation~\eqref{eq:n}.
\begin{equation}
N_{i}=\left \lceil \frac{SI\times\rho_{i}}{L_{i}} \right \rceil,
\label{eq:n}
\end{equation}
where $L_{i}$ is the nominal MSDU length for the $i^{th}$ QSTA.

Then the TXOP duration of the particular station, $TXOP_{i}$, is calculated as the maximum of the time required to transmit $N_{i}$ MSDU or the time to transmit one maximum MSDU at the physical rate $R_{i}$, as stated in Equation~\eqref{eq:txop}.
\begin{equation}
TXOP_{i}=max\left (\frac{N_{i} \times L_{i}}{R_{i}} + O, \frac{M}{R_{i}} + O \right)
\label{eq:txop}
\end{equation}
where $M$ is the Maximum MSDU Size and $O$ is the overhead including MAC and PHY headers, inter-frame spaces (IFSs), and the acknowledgment and poll frames overheads.
\subsubsection{Admission Control}
The ACU manages the TSs admission while maintaining the QoS of the already admitted ones. When a new TS demands an admission, the ACU First obtains a new $SI$ as shown in the previous step and calculates the number of MSDUs expected to arrive at the new $SI$ using Equation~\eqref{eq:n}. Then it calculates the $TXOP_{i}$ for the particular TS using Equation~\eqref{eq:txop}. Finally, ACU only admits the TS if the following inequality satisfied.
\begin{equation}
\label{eq:ACU}
\frac{TXOP_{n+1}}{SI}+\sum_{i=1}^{n} \frac{TXOP_{i}}{SI}\leq \frac{T- T_{CP}}{T}
\end{equation}
where $n$ is the number of currently admitted TSs, so that ($n+1$) represents the incoming TS, $T$ is the beacon interval and $T_{CP}$ is the duration reserved for EDCA. The HC sends an acceptance message (ADDTS-Response) to the requested QSTA if the condition in Equation~\eqref{eq:ACU} is true or send a rejection message otherwise. The accepted TS will be added to the polling list of the HC.
\subsection{Variable Bit Rate MPEG--4 Video Traffic}
\label{sec:VBRTraffic}
MPEG--4 is an efficient video encoding covering a wide domain of bit rate coding ranging from low-bit-rate for wireless transmission up to higher quality beyond high-definition television (HDTV) \cite{Fitzek2001}. For this reason, MPEG--4 video coding has become from among the prominent video traffics in the internet nowadays. 

In fact, MPEG--4 videos are encoded using different compression ratios which yield to produce different levels of quality. Higher compression level generates lower-quality video with smaller mean frame sizes and smaller mean bit rate and vice versa. This variability in the compression level is adequate to transmit the video packets over the limited wireless network resources such as low bit rate. Table \ref{tab:traceFragHigh} displays excerpts of video trace file of Jurassic Park 1 movie \cite{Fitzek2000} encoded using MPEG--4 at high quality.
In MPEG--4 video coding, successive pictures of the coded video stream compose a Group of Picture (GoP) which identifies how the intra- (I-frame) and inter-frames (P- and B- frames) are ordered.
Here, we display one GoP of encoded Jurassic Park 1 video which consists the pattern IBBPBBPBBPBB. One can notice that the frame are not sequenced chronologically, yet rather it is ordered according to the display time instead.
\begin{table}
\centering
\caption {A fragment of Jurassic Park 1 trace file encoded using MPEG--4 at high bit rate}
\begin{tabular}{cccc}
\hline
Frame sequence & Frame type & Frame period (ms) & Frame size (byte)
\\ \hline
527 & I & 21120 & 8124 \\
528 & B & 21040 & 6442 \\
529 & B & 21080 & 6237 \\
530 & P & 21240 & 7581 \\
531 & B & 21160 & 6184 \\
532 & B & 21200 & 6173 \\
533 & P & 21360 & 7482 \\
534 & B & 21280 & 6331 \\
535 & B & 21320 & 6567 \\
536 & P & 21480 & 7130 \\
537 & B & 21400 & 6410 \\
538 & B & 21440 & 6223 \\ \hline
\end{tabular}
\label{tab:traceFragHigh}
\end{table}
As it is mentioned above, HC schedules QSTAs with respect to the negotiated TSPEC parameters that represent the mean characteristics of their traffics. Basically, the weakness of HCCA in supporting VBR traffic is because of the lack of information about the abrupt changing in the video traffic profile during the time, the traffic burstiness. In the case of the uplink traffic, from QSTA to QAP, and for the transmission of prerecorded upnlik video traffics, it will be beneficial to send feedback information about the changing in the video profile to accommodate the fast fluctuation of the traffic.

\hspace{1mm}Several approaches such as \cite{Lee2009, Jansang2011, Cecchetti2012, cecchettielAL2012, ruscelli2013} have been presented in the literature attempting to remedy the deficiency of the HCCA reference scheduler in supporting QoS for VBR traffics. However, these enhancements are still not sufficient to cope with the fast fluctuating nature of highly compressed video applications since the QSTAs are scheduled according to an estimation about the uplink TSs characteristic which may be far from the real traffics.

\section{Adaptive TXOP Scheduling Algorithm}
\label{sec:ATAV}
HCCA scheduler computes TXOP durations by estimating the amount of data expected to be transmitted by the QSTA during SI. This estimation is based on the TSPEC negotiated with HC which considers the mean characteristics of the traffic. The proposed scheduling mechanism described in this section is referred to as “Adaptive Transmission Opportunity” as it adapts TXOP duration based on the feedback of the next frame packet size reported by QSTAs. The proposed mechanism gives an actual TXOP needed by stations and ensures that the end-to-end delay is minimized without jeopardizing the channel bandwidth. The scheduling parameters along with the scheduling operation are described below.

\subsection{Scheduling Parameters}
As the proposed mechanism operates based on the feedback information about the next frame size, the HCCA scheduler will change  some of the parameters in Equation~\eqref{eq:txop} upon receiving a feedback from QSTA. Herein a description of these parameters:

\subsubsection{Number of MSDUs Received in SI ($N_{i}$)} using Equation~\eqref{eq:n}, the reference scheduler calculates the expected number of received packets every SI based on mean TSPEC parameters at the traffic setup phase. In our mechanism, this parameter is set to 1 as only one packet is expected to be generated at the QSTA every SI.
\label{enm:N_i}
\subsubsection{Mean Size of MSDU ($L_{i}$)} the HC updates $L_{i}$ in Equation~\eqref{eq:txop} with regards to the information piggybacked with each packet received from a QSTA. In fact, this is the major part of the proposed mechanism in which the TXOP duration given to a QSTA is calculated dynamically to accommodate the actual packet size to be received at the QAP.
\label{enm:L_i}
\subsection{Adaptive TXOP Mechanism Operation}
In this mechanism, the exact MSDU size of the next frame of the uplink stream is obtained from the application layer through cross layering. For this reason we call it Adaptive TXOP assignment. This information is transmitted with each packet to the QAP carrying the next frame size. Upon each data frame reception, the HC recalculates the TXOP duration to be granted for the specific station in the next SI so as to adapt to the fast varying in VBR video traffic and consequently minimize the packet end-to-end delay and conserve the channel utilization. In this section, we present the description of TXOP operation at both QSTAs and the QAP.
\subsubsection{Operation at the station}
At the QSTA, information about the next MSDU frame size is obtained from the application layer via cross-layering. This information is carried in the Queue  Size (QS) field introduced by IEEE 802.11 standard \cite{IEEEStandard2007} which is a part of the QoS Control field of the QoS data frame. The QS field is exploited in this mechanism for sending information about the next MSDU frame size to the QAP.
\subsubsection{Operation at the access point} 
After the traffic setup phase, the QAP transmits the first poll frame granting the QSTA a TXOP duration. The station will accordingly transmit the first packet of its traffic to the QAP. Note that the inter-arrival time between encoded video traffic frames is a multiple of a fixed interval (typically 40 ms) depends on the encoding parameters. That is to say, it is expected to receive only one packet at a multiple of a designated interval. Details about the operation of our mechanism at QAP is reported in Algorithm~\ref{algo01}.
\begin{algorithm}
\caption{Adaptive TXOP Mechanism Pseudo Code}\label{algo01}
\textbf{INPUT:}

$Stations$, a list of $N$ station in the polling list of the HC\;
$Sizes$, a list of next packet sizes for each $station_{i}$ in the $Stations$ list where $i=1..N$\;
AT THE EVENT OF RECEIVING A DATA PACKET FROM $station_{i}$

save the packet size of $station_{i}$ in $Sizes$

\For{each CAP}{
\While{$Stations$ is not null}{
\uIf{no data packet received from $station_{i}$}{
$L_{i} \leftarrow MSDU_{i}$

obtain $TXOP_{i}$ from Equation~\eqref{eq:txop}
}
\Else{
obtain $TXOP_{i}$ from Equation~\eqref{eq:atav}
}
Poll $station_{i}$
}
}
\end{algorithm}\DecMargin{1em}

At the beginning of each CAP, HC will go through the $stations_{i}$ list and compute $TXOP_{i}$ for the $station_{i}$ according to one the two cases: one case is when a data packet is received from the $station_{i}$ in the previous CAP/SI period, the MSDU size ($Size_{i}$) of the next frame is obtained from the QS field of IEEE 802.11e MAC header. Then, a $TXOP_{i}$ of $QSTA_{i}$ is calculated using Equation~\eqref{eq:atav}. 
\begin{equation}
TXOP_{i}= \frac{Size_{i}}{R_{i}} + O
\label{eq:atav}
\end{equation}
The other case when no data packet is received due to loss, the QAP will use the Equation~\eqref{eq:txop} of  HCCA scheduler to compute the $TXOP_{i}$. It is worth noting that at the first CAP of any TS, the TXOP is calculated based on Equation~\eqref{eq:txop} because no information about the next packet size has been reported yet.

\section{Performance Evaluation}
\label{sec:evaluation}
To measure the performance of the proposed mechanism, we have used a network simulation tool. The simulation environment setup, and video traffic used as uplink traffics is described in details in this section. The performance of our mechanism is compared against the reference design of the HCCA. The results of end-to-end delay and throughput are also discussed.
\subsection{Simulation Setup}
The software implementation of the proposed mechanism has been developed on a network simulator (\textit{ns-2}) \cite{NS2} version 2.27. The HCCA implementation framework \textit{ns2hcca} \cite{cicconetti2005} has been patched to provide the controlled access mode of IEEE 802.11e functions, HCCA. The \textit{ns-2} traffic trace agent is used for video stream generation.

\begin{figure}[hbtp]
\centering
\includegraphics[scale=0.43]{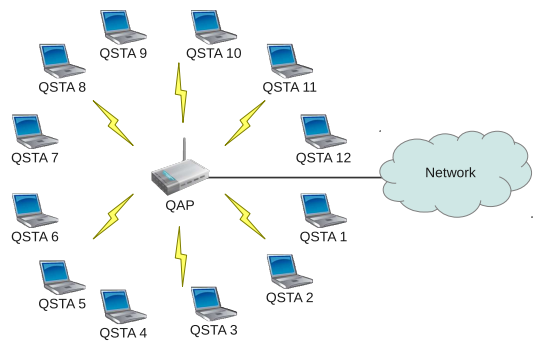}
\caption{Network Topology}
\label{fig:topology}
\end{figure}

A star topology has been used for constructing the simulation scenario to form an infrastructure network with one QAP surrounded by varying number of QSTAs ranging from 1 to 12. All QSTAs were distributed uniformly around the QAP with a radius of 10 meters as shown in Fig. \ref{fig:topology}. The stations were placed within the QAP coverage area, in the same basic service set BSS, and the wireless channel is assumed to be ideal. Since we focus on HCCA performance measurement, all the stations operate only on the contention-free mode by setting $T_{CP}$ in Equation~\eqref{eq:ACU} to zero.
QAP is the sink receiver, while all stations are the video sources each send only an uplink video traffic as only one flow per station is supported in \textit{ns2hcca} patch. Therefore, for simulating concurrent video streams multiple stations are added each with one flow. In order to leave an ample time for initialization, stations start their transmission after 20 (sec) from the start of the simulation time and last till the simulation end. Wireless channel assumed to be an error-free, and no admission control used for the sake of investigating the maximum scheduling capability of each examined algorithm under heavy traffic conditions. Simulation parameters are summarized in Table \ref{tab:SimPars}.
\begin{table}
    \caption {Simulation Parameters}
    \centering
    \begin{tabular}[width=\linewidt]{p{6cm}p{2.2cm}}
    \hline
    Parameter			& Value			\\ \hline
    Simulation time		& 500 sec		\\
    MAC layer			& IEEE 802.11e	\\
    SIFS				& 10 $\mu s$	\\
    PIFS				& 30 $\mu s$	\\
    Slot time			& 20 $\mu s$	\\
    Preamble length		& 144 bits		\\
    PLCP header length	& 48 bits		\\
    MAC header			& 36 bytes		\\
    Data rate			& 11 Mbps		\\
    Basic rate			& 1 Mbps		\\ \hline
    \end{tabular}
    \label{tab:SimPars}
\end{table}

For evaluating the performance of our mechanism against the reference scheduler of HCCA, Jurassic Park 1 video sequence trace encoded using MPEG--4 was chosen from a publicly available library for video traces \cite{Fitzek2001}. We tested the proposed mechanism with three levels of comparison low, medium and high which result in different fluctuating in packet size. Table \ref{tab:traceStats} demonstrates some statistics of the examined traces. TSPEC parameters used for each video traffic is shown in Table \ref{tab:VideoParas} with regards to video QoS requirements.
\begin{table}
\centering
\caption {Frame Statistics of MPEG--4 Jurassic Park 1 Movie Trace File}
    \begin{tabular}{lllll}
\hline
Parameter				& low quality	& medium quality& high quality	\\
\hline
Comp. ratio (YUV:MP4)	& 49.46			& 28.4			& 9.92			\\
Mean size (byte)		& 770			& 1300			& 3800			\\
Mean bit rate (bit/sec)	& 1.5e+05		& 2.7e+05		& 7.7e+05		\\
Peak bit rate (bit/sec)	& 1.6e+06		& 1.7e+06		& 3.3e+06		\\
Peak/Mean of bit rate	& 10.61			& 6.36			& 4.37			\\ \hline 
\end{tabular}
\label{tab:traceStats}
\end{table}
\begin{table}
\centering
\caption {Traffic Parameters for Jurassic Park 1 Video}
    \begin{tabular}[width=\linewidth]{lllll}
    \hline
    Parameter	& unit		& low quality	& medium quality	& high quality	\\ \hline
    $L$			& bytes		& 7.7e+02		& 1.3e+03			& 3.8e+03 		\\
    $M$			& bytes		& 8154			& 8511				& 16745 		\\
    $\rho$		& bit/sec	& 1.5e+05		& 2.7e+05			& 7.7e+05		\\
    $D$			& sec		& 0.08			& 0.08				& 0.08			\\
    $R$			& Mbps		& 11			& 11				& 11			\\
    $MSI$		& sec		& 0.04			& 0.04				& 0.04			\\ \hline
    \end{tabular}
\label{tab:VideoParas}
\end{table}
\subsection{Results and Discussion}
Simulations have been carried out to exhibit the performance of the examined mechanisms using different variability level of the same videos. Since the main objective is to achieve superior QoS support by accurately granting TXOP to the station so that it fits its need, packet end-to-end delay of the uplink traffics has been measured which considered as one of the significant metrics to evaluate a QoS support of video streams. To validate the behavior of the examined mechanisms, the measurements is done for an increasing number of TSs. The system throughput was also investigated to verify that the improvement in the delay is achieved without jeopardizing the wireless channel efficiency.

\subsubsection{End-to-End Delay Analysis}
\begin{figure}[t]
\centering
\subfigure[Low-Quality Encoded Jurassic Park 1 Video]{
\label{fig:e2eLowDly}
\includegraphics[scale=0.49]{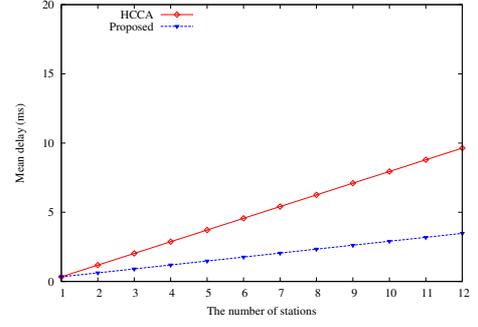}
}
\subfigure[Medium-Quality Encoded Jurassic Park 1 Video]{
\label{fig:e2eMedDly}
\includegraphics[scale=0.49]{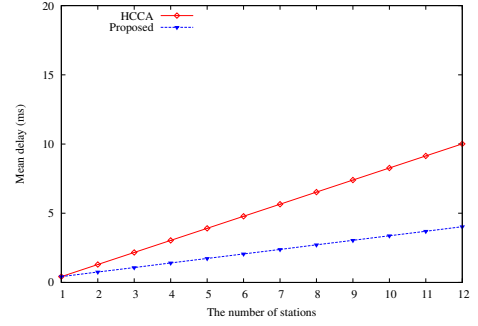}
}
\subfigure[High-Quality Encoded Jurassic Park 1 Video]{
\label{fig:e2eHighDly}
\includegraphics[scale=0.49]{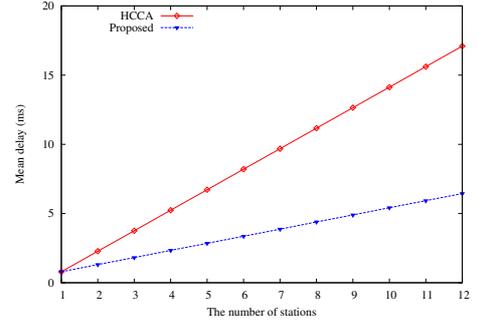}
}
\caption{Packet Mean End-to-End Delay as a Function of Number of Stations}
\label{fig:accessDly}
\end{figure}

The end-to-end delay is defined as the time elapsed from the generation of the packet at the source QSTA application layer until it has been received at the QAP which is expressed in Equation~\eqref{eq:meanDelay}.
\begin{eqnarray}
\label{eq:meanDelay}
e2eDelay  = \frac { \sum_{i=1}^{N} ( R_{i}-G_{i})  } {N},
\end{eqnarray}
where $G_{i}$ is the generation time of packet $i$ at the source QSTA, $R_{i}$ is the receiving time of the particular packet at the MAC layer of the QAP and $N$ is the total number of packets for all flows in the system.The end-to-end delay has been measured for the three video types to study the efficiency of  both HCCA and our mechanisms with different traffic variability. Fig. \ref{fig:e2eLowDly}, \ref{fig:e2eMedDly} and \ref{fig:e2eHighDly} depict the delay experienced by data packets for the low-, medium- and high-quality video, respectively. One can notice that the end-to-end delay boosts with the increase of the packet size, higher quality exhibit higher end-to-end delay and vice versa. Furthermore, the delay improvement in our mechanism is justified by the accurate calculation of the TXOP. Unlike the HCCA scheduler that only relies on the mean traffic characteristic which is not reflecting the actual traffic behavior.

\subsubsection{Throughput Analysis}
\begin{figure}[t]
\centering
\subfigure[Low-Quality Encoded Jurassic Park 1 Video]{
\label{fig:thrp1}
\includegraphics[scale=0.49]{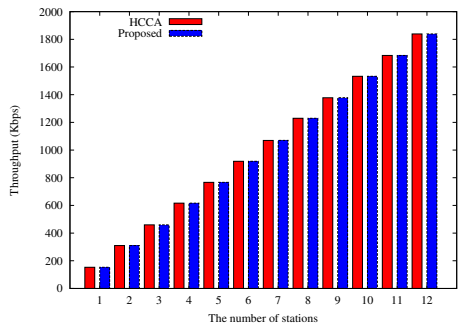}
}
\subfigure[Medium-Quality Encoded Jurassic Park 1 Video]{
\label{fig:thrp2}
\includegraphics[scale=0.49]{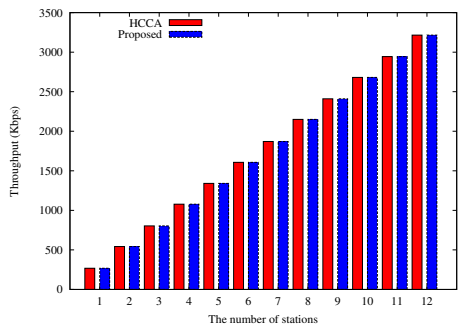}
}
\subfigure[High-Quality Encoded Jurassic Park 1 Video]{
\label{fig:thrp3}
\includegraphics[scale=0.49]{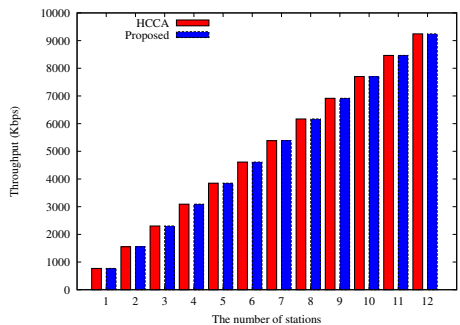}
}
\caption{Aggregate Throughput as a Function of Varying Number of Stations}
\end{figure}

The aggregate throughput of the examined mechanisms has been investigated as a function of the number of stations. This is to verify that our mechanism is efficient in supporting QoS for VBR traffics which maintaining the utilization of the wireless channel. The aggregate throughput is calculated in Equation~\eqref{eq:throughput}.
\begin{eqnarray}
\label{eq:throughput}
AggregateThrp  = \frac { \sum_{i=1}^{N} ( Size_{i})  } {time},
\end{eqnarray}
where $Size_{i}$ is the received packet size at the QAP, $time$ is the simulation time and $N$ is the total number of the received packets at QAP during the simulation time. Fig. \ref{fig:thrp1},  \ref{fig:thrp2} and \ref{fig:thrp3} depict the aggregate throughput with increasing the network load for the low-, medium- and high-quality Jurassic Park 1 videos,  respectively. The results show that the throughput is the same as that achieved by the HCCA scheduling mechanism. This implies that our approach enhanced the end-to-delay without jeopardizing the wireless channel bandwidth.

\section{Conclusion}
\label{sec:conclusion}
This paper proposed a novel scheduling mechanism to support prerecorded VBR video stream transmission in IEEE 802.11e WLANs. This mechanism adaptively assigns TXOP to station based on feedback information about the next frame size with each packet sent of uplink traffic. Accordingly, HC is able to poll stations with regard to their actual need to prevent stations from receiving excessive TXOP which results in a noticeable increase in the end-to-end delay. Simulation results reveal the efficiency of the proposed mechanism over the HCCA scheduler in terms of minimizing the end-to-end delay while maintaining the system throughput.

\section*{Acknowledgment}
This work was supported by the Malaysian Ministry of High Education under the Fundamental Research Grant Scheme, FRGS/02/01/12/1143/FR.


\bibliographystyle{IEEEtran}
\bibliography{atav}

%



\end{document}